\documentclass[preprint]{revtex4}
\usepackage[latin1]{inputenc}
\usepackage{graphicx,epstopdf}
\usepackage{amsmath}
\usepackage{amssymb}
\usepackage{subfigure}
\usepackage{epsfig}
\usepackage{bigints}
\usepackage[usenames]{color}

\begin{document}

\title{Nonlinear waves in magnetized quark matter and the
reduced Ostrovsky equation}

\author{D. A. Foga\c{c}a, S. M. Sanches Jr. and F. S. Navarra}

\address{Instituto de F\'isica, Universidade de S\~ao Paulo,
Rua do Mat\~ao Travessa R, 187, 05508-090 S\~ao Paulo, SP, Brazil}

\begin{abstract}

We study nonlinear waves in a nonrelativistic ideal and cold quark gluon
plasma immersed in a strong uniform magnetic field. In the context of
nonrelativistic hydrodynamics with an external magnetic field we derive a
nonlinear wave equation for baryon density perturbations, which can be
written as a reduced Ostrovsky equation. We find analytical solutions and
identify the effects of the magnetic field.
\end{abstract}


\maketitle

\section{Introduction}

There is a strong evidence that quark gluon plasma (QGP) has been observed
in heavy ion collisions at RHIC and at LHC \cite{qgp1, qgp2}.
Deconfined quark matter may also exist in the core of compact stars
\cite{qgp3}.  Waves may be formed in the  QGP \cite{w1,w2}. In heavy
ion collisions
waves may be produced, for example, by  fluctuations in baryon number, energy
density or temperature caused by  inhomogeneous initial conditions \cite{w2}.

In order to study waves, it is very often assumed that they  represent
small perturbations in a fluid and hence one can linearize the equations of
hydrodynamics and find their solutions, which are linear waves.
Alternatively, instead of linearization we may use another
procedure, called Reductive Perturbation Method (RPM) \cite{rpm},  which
preserves the nonlinearity of the original equations. This leads to
nonlinear differential equations, whose solution describe
nonlinear waves, such as solitons. In a series of works \cite{werev,weset}
we studied the existence and properties of nonlinear waves in
hadronic matter and in a quark gluon plasma as well.

The existence and effects of a magnetic field in quark stars has been
studied since long time ago \cite{mag1} and became a hot topic in our
days. In a different context, about ten years ago \cite{mag2} it was
realized that a very strong magnetic field might be produced in
relativistic heavy ion collisions and it might have some effect on the
quark gluon plasma phase. A natural question is then: what it
the effect of the magnetic field on the  waves propagating through the QGP ?

In a previous work \cite{we17} we studied the conditions for an ideal, cold
and magnetized quark gluon plasma (QGP) to support stable and causal
perturbations.  These perturbations were considered in the linear approach
and the QGP was treated with nonrelativistic hydrodynamics. We have
derived the dispersion relation for density and velocity perturbations. The
magnetic field was included both in the equation of state and in the
equations of motion, where the term of the Lorentz force was considered.
We have used three equations of state: a generic non-relativistic one, the
MIT bag model EOS (for weak and strong magnetic field) and the mQCD EOS. The
anisotropy effects caused by the B field were also manifest in the parallel
and perpendicular sound speeds. We found that the existence of a strong
magnetic field does not lead to instabilities in the velocity and density
waves. Moreover, in most of the considered cases the propagation of these
waves was found to respect causality. However causality might be violated
in the strong field  regime. The onset of causality violation might happen
at very large densities and/or large values of the wave length (small values
of the wavenumber $k$).
The magnetic field changes the pressure, the  energy density and the speed
of sound. It also changes the equations of hydrodynamics.
One of the conclusions of Ref. \cite{we17} is that  the  changes in
hydrodynamics are by far more important than the changes in the equation of
state.

In the present study we extend our previous work to the case of nonlinear
waves.  We will investigate the effects of a strong and uniform magnetic
field on nonlinear baryon density perturbations in an ideal and magnetized
quark gluon plasma.  We consider the magnetic field in the EOS and also in
the Euler equation,
which requires special attention in the RPM \cite{rpm} formalism.
Our study could be applied to the deconfined cold quark matter in compact
stars and to the cold quark gluon plasma formed in heavy ion collisions at
intermediate energies at FAIR \cite{fair} or NICA \cite{nica}.
We go beyond the linear
approach used in \cite{we17} and improve the nonlinear treatment used in
\cite{w2}, now including the strong magnetic field effects.

Some work along this line was already published in \cite{azam}, where the
authors concluded that increasing the magnetic
field leads to a reduction in the amplitude of the nonlinear waves.
More recently \cite{javi17}, perturbations in a cold QGP were studied with
nonrelativistic hydrodynamics with magnetic field effects in a nonlinear approach.
Solitonic density waves were found as solutions of a
modified  nonlinear  Schrodinger equation.   The magnetic field was  found to
increase the phase speed of the soliton and to reduce its width.
We will discuss below the differences between our study and the above mentioned
works.

\section{Nonrelativistic hydrodynamics}

We start from the nonrelativistic Euler equation \cite{land} with an external
uniform magnetic field. The same magnetic field affects the thermodynamical
quantities appearing in the equation of state, as  in \cite{we17}.  The
magnetic field of intensity $B$ is chosen to be in the $z-$direction and hence
$\vec{B}=B \hat{z}$ . The three fermions species considered are the quarks: up ($u$),
down ($d$) and
strange ($s$) with the following respectively charges $Q_{u}= 2 \, Q_{e}/3$,
$Q_{d}= - \, Q_{e}/3$  and $Q_{s}= - \, Q_{e}/3$,
where $Q_{e}=0.08542$ is the absolute value of the electron charge in
natural units \cite{glend}. Because of the external  magnetic field, particles with different
charges  may assume different trajectories \cite{azam,multif} and this
justifies  the use of the multi-fluid approach \cite{azam,multif,we17}.
Throughout this work, we employ natural units ($\hbar=c=1$) and the metric used is
$g^{\mu\nu}=\textrm{diag}(+,-,-,-)$.

Starting from the hydrodynamics equations discussed in \cite{we17}, and
The Euler equation for the quark of flavor $f$ (f=u,d,s) reads:
\begin{equation}
{\rho_{m\,f}}\Bigg[{\frac{\partial \vec{v_f}}{\partial t}} +
(\vec{v_f}\cdot \vec{\nabla}) \vec{v_f}\Bigg]=
-\vec{\nabla}p
+{\rho_{c\,f}}\Big(\vec{v_f} \times \vec{B} \Big)
\label{nsgeralmag}
\end{equation}
where ${\rho_{m\,f}}$ is the  quark mass density.  The
charge density of the quark flavor $f$ is
$\rho_{c\,f}$ \cite{azam} and the masses are: $m_{u}=2.2 \, MeV$,
$m_{d}=4.7 \, MeV$, $m_{s}=96 \, MeV$
and $m_{e}=0.5 \, MeV$ \cite{pdg}.

The continuity equation for the  mass density $\rho_{m\,f}$ is\cite{land}:
\begin{equation}
{\frac{\partial \rho_{m\,f}}{\partial t}} + \vec{\nabla}  \cdot (\rho_{m\,f} \,
{\vec{v_f}})=0
\label{conteq}
\end{equation}
The relationship between the mass density and the baryon density is
${\rho_{m}}_f=3m_{f} \,\, {\rho_{B}}_{f}$ \cite{we17}. The  charge density for
each quark is given
by ${\rho_{c}}_{u}=2Q_{e}\,{\rho_{B}}_{u}$\, ,
\, ${\rho_{c}}_{d}=-Q_{e}\,{\rho_{B}}_{d}$ \,\, and \,\,
${\rho_{c}}_{s}=-Q_{e}\,{\rho_{B}}_{s}$ . In general we write
${\rho_{c}}_{f}=3\,{Q_{f}} \, {\rho_{B}}_{f}$ for each quark $f$.

\section{Equation of state}

In general, the equation of state (EOS) of the quark gluon plasma can be
written as a  relation between pressure $p$ and energy density
$\epsilon$: $p = {c_s}^{2}\epsilon$, where $c_s$ is the speed of sound.
As previously studied in \cite{we17,we16,soundes}, when the fluid is immersed
in an external uniform magnetic field, the pressure splits into
a parallel (with respect to the direction of the external field),
$p_{\parallel}$, and a perpendicular  component, $p_{\perp}$.
We have thus a parallel (${c_s}_{\parallel}$) and a perpendicular
(${c_s}_{\perp}$) speed of sound,  given by \cite{we17,we16,soundes}:
\begin{equation}
{({c_{s}}_{\parallel})}^{2}={\frac{\partial p_{\parallel}}{\partial \varepsilon}}
\hspace{1.0cm} \textrm{and} \hspace{1.0cm}
{({c_{s}}_{\perp})}^{2}={\frac{\partial p_{\perp}}{\partial \varepsilon}}
\label{soundes}
\end{equation}
and hence $p_{\parallel} \approx {({c_{s}}_{\parallel})}^{2} \, \varepsilon$ \,
and \, $p_{\perp} \approx {({c_{s}}_{\perp})}^{2} \, \varepsilon$\, .
The pressure gradient can then be written as:
\begin{equation}
\vec{\nabla}{p}=\Bigg({\frac{\partial {p}_{\perp}}{\partial x}}\,,\,
{\frac{\partial {p}_{\perp}}{\partial y}}\,,\,{\frac{\partial {p}_{\parallel}}{\partial z}}\Bigg)  
\label{gradpresscartsforquarks}
\end{equation}

\subsection{The nonrelativistic equation of state}

As in \cite{we17}, we take here the limit \cite{w2}: $\varepsilon \cong {\rho_{m}}$.
Since ${\rho_{m}}=3m_{f}\,{\rho_{B}}_{f}$ and remembering that the pressure is anisotropic, the
pressure gradient (\ref{gradpresscartsforquarks})  for the quark of flavor f is given by \cite{we17}:
\begin{equation}
\vec{\nabla} p = 3m_{f}\Bigg({({c_{s}}_{\perp})}^{2}\,{\frac{\partial{{\rho_{B}}_{f}}}{\partial x}}\,,\,
{({c_{s}}_{\perp})}^{2}\,{\frac{\partial{{\rho_{B}}_{f}}}{\partial y}}\,,\,
{({c_{s}}_{\parallel})}^{2}\,{\frac{\partial{{\rho_{B}}_{f}}}{\partial z}}\Bigg)  
\label{gradespresses}
\end{equation}

\subsection{The improved  MIT equation of state}

The EOS which we call mQCD  was derived in  \cite{we11} and used in \cite{we16}  and also in
\cite{we17}.  The energy density ($\varepsilon$), the parallel pressure ($p_{f\,\parallel}$) and the
perpendicular pressure ($p_{f\,\perp}$), are given respectively by
\cite{we16,we17}:
\begin{equation}
\varepsilon={\frac{27{g_{h}}^{2}}{16{m_{G}}^{2}}}\,({{\rho_{B}}})^{2}
+{\mathcal{B}}_{QCD}+{\frac{B^{2}}{8\pi}}
+\sum_{f=u}^{d,s}{\frac{|Q_{f}|B}{2\pi^{2}}}\sum_{n=0}^{n^{f}_{max}} 3(2-\delta_{n0})
\int_{0}^{k^{f}_{z,F}} dk_{z}\sqrt{m_{f}^{2}+k_{z}^{2}+2n|Q_{f}|B} \, ,
\label{epsilontempzeromagon}
\end{equation}
\begin{equation}
p_{\parallel}={\frac{27{g_{h}}^{2}}{16{m_{G}}^{2}}}\,({{\rho_{B}}})^{2}
-{\mathcal{B}}_{QCD}-{\frac{B^{2}}{8\pi}}
+\sum_{f=u}^{d,s}{\frac{|Q_{f}|B}{2\pi^{2}}}  \sum_{n=0}^{n^{f}_{max}} 3(2-\delta_{n0})
\int_{0}^{k^{f}_{z,F}} dk_{z} \,{\frac{{k_{z}}^{2}}{\sqrt{m_{f}^{2}+k_{z}^{2}+2n|Q_{f}|B}}}
\label{parallelpressuremagon}
\end{equation}
and
\begin{equation}
p_{\perp}={\frac{27{g_{h}}^{2}}{16{m_{G}}^{2}}}\,({{\rho_{B}}})^{2}
-{\mathcal{B}}_{QCD}+{\frac{B^{2}}{8\pi}}
+\sum_{f=u}^{d,s}{\frac{|Q_{f}|^{2}B^{2}}{2\pi^{2}}}  \sum_{n=0}^{n^{f}_{max}} 3(2-\delta_{n0})
n \int_{0}^{k^{f}_{z,F}}   {\frac{dk_{z}}{\sqrt{m_{f}^{2}+k_{z}^{2}+2n|Q_{f}|B}}}  
\label{perppressuremagon}
\end{equation}
The baryon density ($\rho_{B}$) is given by \cite{we16,we17}:
\begin{equation}
\rho_{B}=\sum_{f=u}^{d,s}\,{\frac{|Q_{f}|B}{2\pi^{2}}} \,
\sum_{n=0}^{n^{f}_{max}}(2-\delta_{n0}) \, \sqrt{{\nu_{f}}^{2}-m_{f}^{2}-2n|Q_{f}|B}
\hspace{0.7cm} \textrm{with} \hspace{0.5cm}
n\leq n^{f}_{max}=int\Bigg[ {\frac{{\nu_{f}}^{2}-m_{f}^{2}}{2|Q_{f}|B}} \Bigg]
\label{magbaryondens}
\end{equation}
where $ {\it{int}}[a]$ denotes the integer part of $a$ and ${\nu_{f}}$ is the chemical
potential for the quark $f$.  As in \cite{we16} we  define $\xi \equiv g_{h}/m_{G}$.
Choosing $\xi=0$ we recover the MIT EOS. For a given magnetic field intensity, we choose
the values for the chemical potentials ${\nu_{f}}$ which determine the density $\rho_{B}$.
We also choose the other parameters: $\xi$ and ${\mathcal{B}}_{QCD}$.
In this case the  pressure gradient (\ref{gradpresscartsforquarks}) becomes
\begin{equation}
\vec{\nabla} p = \Bigg({\frac{27{g_{h}}^{2}}{8{m_{G}}^{2}}}\Bigg)
\Bigg({\rho_{B}}_{f}\,{\frac{\partial{{\rho_{B}}_{f}}}{\partial x}}\,,\,
{\rho_{B}}_{f}\,{\frac{\partial{{\rho_{B}}_{f}}}{\partial y}}\,,\,
{\rho_{B}}_{f}\,{\frac{\partial{{\rho_{B}}_{f}}}{\partial z}}\Bigg)   
\label{mqcdgradespresses}
\end{equation}

\section{Nonlinear waves}

Now we apply the Reductive Perturbation Method (RPM)
\cite{rpm,w2,weset,javi17,azam} to the basic
equations of hydrodynamics (\ref{nsgeralmag}) and (\ref{conteq}) to obtain the nonlinear
wave equations that govern the baryon density perturbations.  The RPM technique goes beyond the
linearization approach and preserves nonlinear terms in the wave equations.
The background density, upon which  small perturbations occur, is defined by $\rho_{0}$, and it
is usually given in terms of the ordinary nuclear matter density $\rho_{N}=0.17\, fm^{-3}$.

According to the RPM technique we rewrite the equations (\ref{nsgeralmag}) changing variables and going
from the $(x,y,z,t)$ space to the  $(X,Y,Z,T)$ space using the ``stretched coordinates'' defined by
\cite{w2,weset}: $X=\sigma^{1/2}(x-{c_{s\,\perp}}\,t)\hspace{0.1cm}, \hspace{0.2cm}Y=\sigma \,
y\hspace{0.1cm}, \hspace{0.2cm} Z=\sigma \,z \hspace{0.5cm} \textrm{and} \hspace{0.5cm}
T=\sigma^{3/2}\,t$.  In our approach, following the RPM algebraic procedure \cite{rpm,weset} we apply
the following transformation to the magnetic field: $B=\sigma\, \tilde{B}$.  In this way, we obtain the equations (\ref{nsgeralmag}) and (\ref{conteq}) in the $(X,Y,Z,T)$ space containing the (small)
parameter $\sigma$, which is the expansion parameter of the dimensionless density and dimensionless
velocities:
\begin{equation}
\hat\rho_{B\,f}(x,y,z,t)={\frac{\rho_{B\,f}(x,y,z,t)}{\rho_{0}}}=1+\sigma {\rho_{f}}_{1}(x,y,z,t)+
\sigma^{2} {\rho_{f}}_{2}(x,y,z,t) + \sigma^{3} {\rho_{f}}_{3}(x,y,z,t)+ \dots  \, ,
\label{roexpa}
\end{equation}
\begin{equation}
{\hat{v}_{f\,x}}(x,y,z,t)={\frac{{v}_{f\,x}(x,y,z,t)}{{c_{s\,\perp}}}}=
\sigma {{v_{{f}\,x}}_{1}}(x,y,z,t)+ \sigma^{2} {{v_{{f}\,x}}_{2}}(x,y,z,t) +
\sigma^{3} {{v_{{f}\,x}}_{3}}(x,y,z,t)+ \dots  \, ,
\label{vxfexpa}
\end{equation}
\begin{equation}
{\hat{v}_{f\,y}}(x,y,z,t)={\frac{{v}_{f\,y}(x,y,z,t)}{{c_{s\,\perp}}}}=
 \sigma^{3/2} {{v_{{f}\,y}}_{1}}(x,y,z,t)+ \sigma^{2} {{v_{{f}\,y}}_{2}}(x,y,z,t) +
\sigma^{5/2} {{v_{{f}\,y}}_{3}}(x,y,z,t)+ \dots
\label{vyfexpa}
\end{equation}
and
\begin{equation}
{\hat{v}_{f\,z}}(x,y,z,t)={\frac{{v}_{f\,z}(x,y,z,t)}{{c_{s\,\parallel}}}}=
\sigma^{3/2} {{v_{{f}\,z}}_{1}}(x,y,z,t)+ \sigma^{2} {{v_{{f}\,z}}_{2}}(x,y,z,t) +
\sigma^{5/2} {{v_{{f}\,z}}_{3}}(x,y,z,t)+ \dots   
\label{vzfexpa}
\end{equation}
Next we use (\ref{roexpa}) to (\ref{vzfexpa}) to rewrite (\ref{nsgeralmag}) and (\ref{conteq}). We
then neglect terms proportional to  $\sigma^{n}$ for $n > 2$ and collect the remaining terms in a
power series of $\sigma$, $\sigma^{3/2}$ and $\sigma^{2}$, solving them in order to obtain an equation
in the $(X,Y,Z,T)$ space.  This equation is finally written back in the usual $(x,y,z,t)$ space,
yielding the nonlinear wave equation for the baryon density perturbation.

The continuity equation (\ref{conteq}) in the RPM gives:
$$
\sigma\Bigg\{-{\frac{\partial{{\rho_{f}}_{1}}}{\partial X}}+{\frac{\partial{{v_{f\,x}}_{1}}}
{\partial X}}\Bigg\}+
\sigma^{2}\Bigg\{-{\frac{\partial{{\rho_{f}}_{2}}}{\partial X}}+{\frac{\partial{{v_{f\,x}}_{2}}}
{\partial X}}+{\frac{1}{({c_{s\,\perp}})}}{\frac{\partial{{\rho_{f}}_{1}}}{\partial T}}+
{\rho_{f}}_{1}{\frac{\partial{{v_{f\,x}}_{1}}}{\partial X}}+
{v_{f\,x}}_{1}{\frac{\partial{{\rho_{f}}_{1}}}{\partial X}}
$$
\begin{equation}
+{\frac{\partial{{v_{f\,y}}_{1}}}{\partial Y}}+\Bigg({\frac{{c_{s\,\parallel}}}{{c_{s\,\perp}}}}\Bigg)
{\frac{\partial{{v_{f\,z}}_{1}}}{\partial Z}}\Bigg\}=0
\label{conteqcartes}
\end{equation}
and the Euler equation (\ref{nsgeralmag}) will be studied in following subsections.

\subsection{Nonrelativistic EOS}

Applying the RPM procedure to Eq. (\ref{nsgeralmag}) and using (\ref{gradespresses}), we obtain
the following set of equations in powers of the $\sigma$ parameter:
\begin{equation}
\sigma\Bigg\{-{\frac{\partial{{v_{f\,x}}_{1}}}{\partial X}}+
{\frac{\partial{{\rho_{f}}_{1}}}{\partial X}}\Bigg\}
$$
$$
+\sigma^{2}\Bigg\{-{\frac{\partial{{v_{{f}\,x}}_{2}}}{\partial X}}
+{\frac{1}{({c_{s\,\perp}})}}{\frac{\partial{{v_{f\,x}}_{1}}}{\partial T}}
+\,{v_{f\,x}}_{1}\,{\frac{\partial{{v_{f\,x}}_{1}}}{\partial X}}
-{\rho_{{f}}}_{1}\,{\frac{\partial{{v_{f\,x}}_{1}}}{\partial X}}
+{\frac{\partial{{\rho_{f}}_{2}}}{\partial X}}
-{\frac{Q_{f}\,\tilde{B}}{m_{f}\,({c_{s\,\perp}})}}{v_{f\,y}}_{1}\Bigg\}=0  \, ,
\label{eulercartex}
\end{equation}
\\
\begin{equation}
\sigma^{3/2}\Bigg\{-{\frac{\partial{{v_{f\,y}}_{1}}}{\partial X}}+
{\frac{\partial{{\rho_{f}}_{1}}}{\partial Y}}+
{\frac{Q_{f}\,\tilde{B}}{m_{f}\,({c_{s\,\perp}})}}{v_{f\,x}}_{1}\Bigg\}
+\sigma^{2}\Bigg\{-{\frac{\partial{{v_{f\,y}}_{2}}}{\partial X}}\Bigg\}=0
\label{eulercartey}
\end{equation}
\\
and
\begin{equation}
\sigma^{3/2}\Bigg\{-{\frac{\partial{{v_{f\,z}}_{1}}}{\partial X}}+
\Bigg({\frac{{c_{s\,\parallel}}}{{c_{s\,\perp}}}}\Bigg){\frac{\partial{{\rho_{f}}_{1}}}{\partial Z}}\Bigg\}
+\sigma^{2}\Bigg\{-{\frac{\partial{{v_{f\,z}}_{2}}}{\partial X}}\Bigg\}=0   
\label{eulercartez}
\end{equation}
Solving the equations (\ref{conteqcartes}) to (\ref{eulercartez}) we arrive at:
\begin{equation}
{\frac{\partial}{\partial X}}\Bigg[{\frac{\partial{{\rho_{f}}_{1}}}{\partial T}}+
({c_{s\,\perp}})\,{\rho_{f}}_{1}{\frac{\partial{{\rho_{f}}_{1}}}{\partial X}}\Bigg]
+{\frac{({c_{s\,\perp}})}{2}}\Bigg[
{\frac{\partial^{2}{{\rho_{f}}_{1}}}{\partial Y^{2}}}
+\Bigg({\frac{{c_{s\,\parallel}}}{{c_{s\,\perp}}}}\Bigg)^{2}{\frac{\partial^{2}
{{\rho_{f}}_{1}}}{\partial Z^{2}}}
\Bigg]={\frac{(Q_{f}\,\tilde{B})^{2}}{2{m_{f}}^{2}\,({c_{s\,\perp}})}}{\rho_{f}}_{1}  
\label{roestretcart}
\end{equation}
Writing (\ref{roestretcart}) back in the cartesian space we obtain the following wave equation:
\begin{equation}
{\frac{\partial}{\partial x}}\Bigg[{\frac{\partial}{\partial t}}{\delta\rho_{B}}_{f}
+({c_{s\,\perp}}){\frac{\partial}{\partial x}}{\delta\rho_{B}}_{f}
+({c_{s\,\perp}}){\delta\rho_{B}}_{f}{\frac{\partial}{\partial x}}{\delta\rho_{B}}_{f}\Bigg]
$$
$$
+{\frac{({c_{s\,\perp}})}{2}}\Bigg[
{\frac{\partial^{2}}{\partial y^{2}}}{\delta\rho_{B}}_{f}
+\Bigg({\frac{{c_{s\,\parallel}}}{{c_{s\,\perp}}}}\Bigg)^{2}
{\frac{\partial^{2}}{\partial z^{2}}}{\delta\rho_{B}}_{f}
\Bigg]=
{\frac{(Q_{f}\,B)^{2}}{2{m_{f}}^{2}\,({c_{s\,\perp}})}}{\delta\rho_{B}}_{f}
\label{roecart}
\end{equation}
where ${\delta\rho_{B}}_{f}\equiv\sigma{\rho_{f}}_{1}$ is the  baryon density perturbation on
the background $\rho_{0}$, as can be seen in (\ref{roexpa}).

Introducing the variable \cite{weset}:
\begin{equation}
\xi=x+y+z
\label{carttransf}
\end{equation}
the equation (\ref{roecart}) becomes:
\begin{equation}
{\frac{\partial}{\partial \xi}}\Bigg\{{\frac{\partial}{\partial t}}{\delta\rho_{B}}_{f}
+\Bigg[{\frac{3}{2}}({c_{s\,\perp}})+{\frac{({c_{s\,\parallel}})^{2}}{2({c_{s\,\perp}})}}\Bigg]
{\frac{\partial}{\partial \xi}}{\delta\rho_{B}}_{f}
+({c_{s\,\perp}}){\delta\rho_{B}}_{f}{\frac{\partial}{\partial \xi}}{\delta\rho_{B}}_{f}\Bigg\}=
{\frac{(Q_{f}\,B)^{2}}{2{m_{f}}^{2}\,({c_{s\,\perp}})}}{\delta\rho_{B}}_{f}  
\label{roecartxitau}
\end{equation}

\subsection{mQCD}

Repeating the steps described in the last subsection and using (\ref{mqcdgradespresses}), we
obtain:
\begin{equation}
\sigma\Bigg\{-{\frac{\partial{{v_{f\,x}}_{1}}}{\partial X}}+
\Bigg({\frac{9\,{g_{h}}^{2}\,{\rho_{0}}}{8\,m_{f}\,{m_{G}}^{2}\,({c_{s\,\perp}})^{2}}}\Bigg)
{\frac{\partial{{\rho_{f}}_{1}}}{\partial X}}\Bigg\}
+\sigma^{2}\Bigg\{-{\frac{\partial{{v_{{f}\,x}}_{2}}}{\partial X}}
+{\frac{1}{({c_{s\,\perp}})}}{\frac{\partial{{v_{f\,x}}_{1}}}{\partial T}}
+{v_{f\,x}}_{1}\,{\frac{\partial{{v_{f\,x}}_{1}}}{\partial X}}
$$
$$
-{\rho_{{f}}}_{1}\,{\frac{\partial{{v_{f\,x}}_{1}}}{\partial X}}
+{\rho_{{f}}}_{1}\,{\frac{\partial{{\rho_{f}}_{1}}}{\partial X}}
+\Bigg({\frac{9\,{g_{h}}^{2}\,{\rho_{0}}}{8\,m_{f}\,{m_{G}}^{2}\,
({c_{s\,\perp}})^{2}}}\Bigg){\frac{\partial{{\rho_{f}}_{2}}}{\partial X}}
-{\frac{Q_{f}\,\tilde{B}}{m_{f}\,({c_{s\,\perp}})}}{v_{f\,y}}_{1}\Bigg\}=0
\label{mqcdeulercartex}  \, ,
\end{equation}
\\
\begin{equation}
\sigma^{3/2}\Bigg\{-{\frac{\partial{{v_{f\,y}}_{1}}}{\partial X}}+
\Bigg({\frac{9\,{g_{h}}^{2}\,{\rho_{0}}}{8\,m_{f}\,{m_{G}}^{2}\,
({c_{s\,\perp}})^{2}}}\Bigg){\frac{\partial{{\rho_{f}}_{1}}}{\partial Y}}+
{\frac{Q_{f}\,\tilde{B}}{m_{f}\,({c_{s\,\perp}})}}{v_{f\,x}}_{1}\Bigg\}
+\sigma^{2}\Bigg\{-{\frac{\partial{{v_{f\,y}}_{2}}}{\partial X}}\Bigg\}=0
\label{mqcdeulercartey}
\end{equation}
\\
and
\begin{equation}
\sigma^{3/2}\Bigg\{-\Bigg({\frac{{c_{s\,\parallel}}}{{c_{s\,\perp}}}}\Bigg)
{\frac{\partial{{v_{f\,z}}_{1}}}{\partial X}}+
\Bigg({\frac{9\,{g_{h}}^{2}\,{\rho_{0}}}{8\,m_{f}\,{m_{G}}^{2}\,
({c_{s\,\perp}})^{2}}}\Bigg){\frac{\partial{{\rho_{f}}_{1}}}{\partial Z}}\Bigg\}
+\sigma^{2}\Bigg\{-\Bigg({\frac{{c_{s\,\parallel}}}{{c_{s\,\perp}}}}\Bigg)
{\frac{\partial{{v_{f\,z}}_{2}}}{\partial X}}\Bigg\}=0
\label{mqcdeulercartez}
\end{equation}
Solving the set of equations (\ref{conteqcartes}) and (\ref{mqcdeulercartex}) to
(\ref{mqcdeulercartez}) we arrive at:
\begin{equation}
{\frac{\partial}{\partial X}}\Bigg[{\frac{\partial{{\rho_{f}}_{1}}}{\partial T}}+
{\frac{3}{2}}({c_{s\,\perp}})\,{\rho_{f}}_{1}{\frac{\partial{{\rho_{f}}_{1}}}{\partial X}}\Bigg]
+{\frac{({c_{s\,\perp}})}{2}}\Bigg(
{\frac{\partial^{2}{{\rho_{f}}_{1}}}{\partial Y^{2}}}
+{\frac{\partial^{2}{{\rho_{f}}_{1}}}{\partial Z^{2}}}
\Bigg)={\frac{(Q_{f}\,\tilde{B})^{2}}{2{m_{f}}^{2}\,({c_{s\,\perp}})}}{\rho_{f}}_{1}
\label{mqcdroestretcart}   
\end{equation}
From the terms of order  $\mathcal{O}(\sigma)$ we obtain the following constraint for the
perpendicular speed of sound:
\begin{equation}
({c_{s\,\perp}})^{2}= {\frac{9\,{g_{h}}^{2}\,{\rho_{0}}}{8\,m_{f}\,{m_{G}}^{2}}}   
\label{csordsigma1lin}
\end{equation}
which coincides with the ``effective sound speed'' ${{\tilde{c}_{s}}}$  obtained in the
linearization approach in \cite{we17}.

Writing (\ref{mqcdroestretcart}) back in cartesian coordinates, we find the following nonlinear wave equation:
\begin{equation}
{\frac{\partial}{\partial x}}\Bigg[{\frac{\partial}{\partial t}}{\delta\rho_{B}}_{f}
+({c_{s\,\perp}}){\frac{\partial}{\partial x}}{\delta\rho_{B}}_{f}
+{\frac{3}{2}}({c_{s\,\perp}}){\delta\rho_{B}}_{f}{\frac{\partial}{\partial x}}{\delta\rho_{B}}_{f}\Bigg]
$$
$$
+{\frac{({c_{s\,\perp}})}{2}}\Bigg(
{\frac{\partial^{2}}{\partial y^{2}}}{\delta\rho_{B}}_{f}
+{\frac{\partial^{2}}{\partial z^{2}}}{\delta\rho_{B}}_{f}
\Bigg)=
{\frac{(Q_{f}\,B)^{2}}{2{m_{f}}^{2}\,({c_{s\,\perp}})}}{\delta\rho_{B}}_{f}
\label{mqcdroecart}
\end{equation}
where again, from (\ref{roexpa}), we have ${\delta\rho_{B}}_{f}\equiv\sigma{\rho_{f}}_{1}$ .
Using (\ref{carttransf}) in (\ref{mqcdroecart}) we find:
\begin{equation}
{\frac{\partial}{\partial \xi}}\Bigg[{\frac{\partial}{\partial t}}{\delta\rho_{B}}_{f}
+2({c_{s\,\perp}}){\frac{\partial}{\partial \xi}}{\delta\rho_{B}}_{f}
+{\frac{3}{2}}({c_{s\,\perp}}){\delta\rho_{B}}_{f}{\frac{\partial}{\partial \xi}}{\delta\rho_{B}}_{f}\Bigg]=
{\frac{(Q_{f}\,B)^{2}}{2{m_{f}}^{2}\,({c_{s\,\perp}})}}{\delta\rho_{B}}_{f}   
\label{mqcdroecartxitau}
\end{equation}

\section{Reduced Ostrovsky Equation (ROE)}

In cartesian coordinates, we derived the ``inhomogeneous three
dimensional breaking wave equations''
given by (\ref{roecart}) and (\ref{mqcdroecart}). By using
(\ref{carttransf}) we transformed these two
equations into (\ref{roecartxitau}) and (\ref{mqcdroecartxitau}),
respectively, where
${\delta\rho_{B}}_{f}(x,y,z,t) \rightarrow {\delta\rho_{B}}_{f}(\xi,t)$.
The equations (\ref{roecartxitau}) and (\ref{mqcdroecartxitau}) can be put
in the form:
\begin{equation}
{\frac{\partial}{\partial \xi}}\Bigg[{\frac{\partial}{\partial t}}{\delta\rho_{B}}_{f}
+{\alpha}\,{\frac{\partial}{\partial \xi}}{\delta\rho_{B}}_{f}
+\beta \,{\delta\rho_{B}}_{f}{\frac{\partial}{\partial \xi}}{\delta\rho_{B}}_{f}\Bigg]=
\Gamma{\delta\rho_{B}}_{f}
\label{roegeral}
\end{equation}
with the nonlinear coefficient $\beta$ and the
velocity of the dispersionless linear wave $\alpha$ defined in
(\ref{roecartxitau}) and (\ref{mqcdroecartxitau}) for each case.
The common dispersion coefficient $\Gamma$ for the two cases is given by
\begin{equation}
\Gamma = {\frac{(Q_{f}\,B)^{2}}{2{m_{f}}^{2}\,({c_{s\,\perp}})}}
\label{disperforced}
\end{equation}
and it comes from the  magnetic field term of the Euler equation
(\ref{nsgeralmag}) for each quark of flavor f.  The magnetic field
effects are also  indirectly present in the coefficients of
(\ref{roegeral}), which come from the equation of state chosen for
the magnetized medium. If the magnetic field were zero, (\ref{roegeral})
would be converted into a breaking wave equation without soliton solutions.
We can then say that the B field allows for localized solitonic solutions
of Eq. (\ref{roegeral}).

Equation (\ref{roegeral}) is known in the literature and it is called
Reduced Ostrovsky equation (ROE) or Ostrovsky-Hunter equation (OHE) when
$\Gamma>0$ \cite{roe}, which is our case.  The ROE is a particular case of the
Ostrovsky equation \cite{ostrov} for a general function $f(\xi,t)$:
\begin{equation}
{\frac{\partial}{\partial \xi}}\Bigg[{\frac{\partial}{\partial t}}f
+ \alpha\,{\frac{\partial}{\partial \xi}}f
+\beta \,f{\frac{\partial}{\partial \xi}}f
+\Pi\,{\frac{\partial^{3}}{\partial \xi^{3}}}f\Bigg]=
\Gamma \,f
\label{ostrovsky}
\end{equation}
when the high-frequency dispersion coefficient $\Pi$ vanishes. Equation
(\ref{ostrovsky})
describes internal waves and weakly nonlinear surface in a rotating ocean
\cite{ostrov}. The equation (\ref{roegeral}) can be solved analytically, as it is shown in the Appendix.

The solution of (\ref{roegeral}) reads:
\begin{equation}
{{\delta\rho_{B}}_{f}}(\xi,t)=-{\frac{6\gamma^{2}\lambda^{2}}
{\beta \Gamma}}sech^{2}
\Big[\lambda\Big(\Omega-\gamma t\Big) \Big]
\label{roesolsexacta}
\end{equation}
where $\lambda$ and $\gamma$ are integration constants. The latter is related to
the propagation speed of the perturbation. Also
\begin{equation}
\xi= x + y + z = \Omega + {\alpha} \, t +\xi_{0}
+{\frac{6 \gamma \lambda}{\beta \Gamma}}\Big\{tanh\Big[\lambda
\Big(\Omega-\gamma t\Big) \Big]
-1\Big\}  
\label{introeagain}
\end{equation}
For a given value of the coordinate $\xi$, we solve the above equation and
find $\Omega$ which is then substituted in (\ref{roesolsexacta}), which
represents a traveling gaussian-looking pulse moving to the right and
preserving its shape.

As it can be seen in (\ref{roesolsexacta}), the amplitude of the density
wave is proportional to $1/\Gamma$ and hence increasing $B$ results in a
decreasing amplitude. Similarly, waves of heavier flavor quarks have larger
amplitudes.  In (\ref{introeagain}) we have
$\alpha = 3({c_{s\,\perp}})/2 + ({c_{s\,\parallel}})^{2}/[2({c_{s\,\perp}})]$
and $\beta={c_{s\,\perp}}$ for the nonrelativistic
EOS. For the mQCD EOS we have
$\alpha = 2({c_{s\,\perp}})$  and  $\beta =  3({c_{s\,\perp}})/2$.

\begin{figure}[h]
\epsfig{file=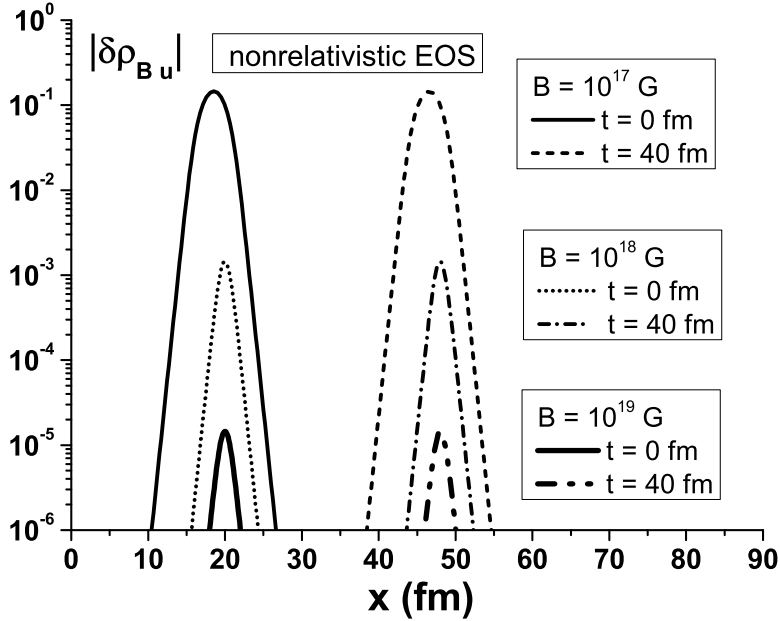,width=135mm}
\caption{Nonrelativistic EOS: solitonic behavior of $|{{\delta\rho_{B}}_{u}}|$
for several times and magnetic field intensities.}
\label{fig1}
\end{figure}

\begin{figure}[h]
\epsfig{file=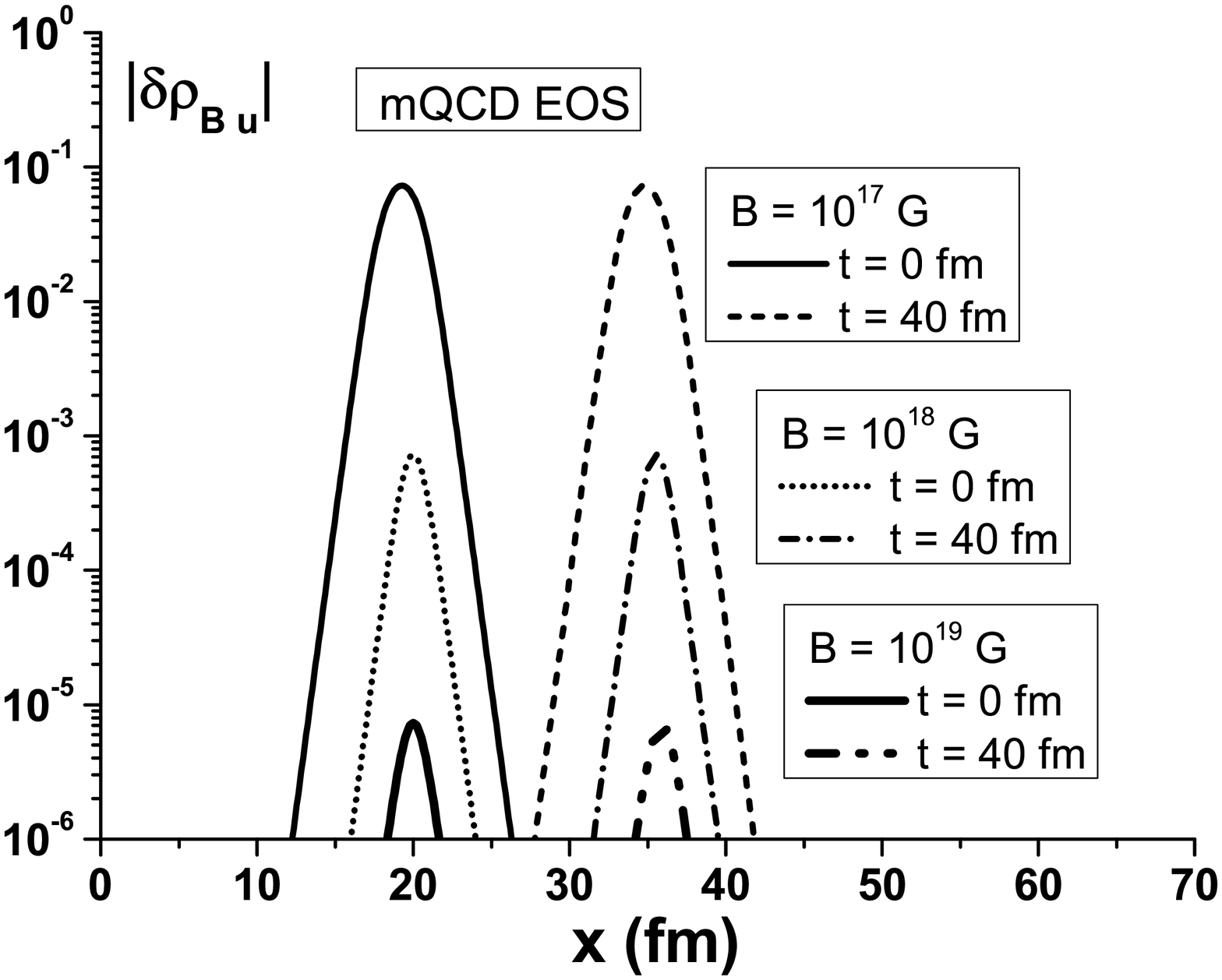,width=135mm}
\caption{mQCD EOS: solitonic behavior of $|{{\delta\rho_{B}}_{u}}|$
for several times and magnetic field intensities.}
\label{fig2}
\end{figure}

To illustrate the solitonic behavior of the rarefaction solution
(\ref{roesolsexacta}), we show in Figs. \ref{fig1} and \ref{fig2}
the perturbation   $|{{\delta\rho_{B}}_{f}}|$
as a function of $x$ for fixed values of $y=0$ and $z=0$ for two values of the
time $t$.  In both cases showed in Figs. \ref{fig1} and \ref{fig2}, we consider the
quark $up$ and  three values of the magnetic field,
that are chosen to satisfy $0<|{{\delta\rho_{B}}_{u}}|<1$ and respect
(\ref{roexpa}) (since ${\delta\rho_{B}}_{u}\equiv\sigma{\rho_{u}}_{1}$).
For magnetic fields $\sim 10^{16} \, G$ or smaller, we obtain
$|{{\delta\rho_{B}}_{u}}|>1$ .

In Fig. \ref{fig1} we show the results obtained with the nonrelativistic
EOS for the parameters
${c_{s\,\perp}}={c_{s\,\parallel}}=0.3$\,, $\xi_{0}=20 \, fm$,
$\lambda=1\, fm^{-1}$ and $\gamma=0.1$ .  The propagation speed of the pulse
is $\alpha+\gamma=0.7$ .

In Fig. \ref{fig2} we show the results obtained with  the mQCD EOS for the
parameters $B_{QCD}=70\,MeV/fm^{3}$, $g_{h}=0.05$, $m_{G}=300\, MeV$,
$\xi_{0}=20 \, fm$, $\lambda=1\, fm^{-1}$ and $\gamma=0.1$ .  The common
chemical potential for all quarks is $\nu_{f}=300 \, MeV$ and for the  chosen
values of the magnetic field we have background densities
$\rho_{0}=2\rho_{N} \sim 2.1\rho_{N}$ $(B=10^{17}\,G \sim 10^{19}\,G)$,
which, with the use of (\ref{csordsigma1lin}) lead to
${c_{s\,\perp}}\cong 0.2$ . The propagation speed of the pulse is
$\alpha+\gamma=0.5$\,, which does not violate causality.

Similar behavior is found when $|{{\delta\rho_{B}}_{u}}|$  is plotted as a
function of the $y$ coordinate (perpendicular to the magnetic field)
and of the  $z$ coordinate (along the magnetic field).

\section{Conclusions}

In this work we focused on nonlinear wave propagation in a cold and
magnetized quark gluon plasma. Including the effects of a strong magnetic
field both in the equation of state and in the basic equations of hydrodynamics,
we derived from the latter a wave equation for a perturbation in the baryon
density.  This wave equation could be identified
as the reduced Ostrovsky equation (ROE), which has a known analytical
solution given by a rarefaction solitonic pulse of the baryon perturbation.
The numerical
analysis and a possible phenomenological application in the context of heavy
ion collisions or in compact stars will be investigated in a future work.
At a qualitative level we can observe
that the most remarkable effect of the magnetic field, as can be seen in the
coefficient $\Gamma$ by (\ref{disperforced}), is to
reduce the wave amplitude. We therefore corroborate and extend the conclusion
found in \cite{azam}.

\vspace{0.5cm}

\section{Appendix}

To establish the integrability of (\ref{roegeral}), we employ the change of
variables developed in \cite{roe}:
\begin{equation}
\xi= \Omega + \beta \int_{-\infty}^{\eta}
\psi(\Omega,\eta') d\eta' \,+\alpha\,\eta +\xi_{0}
\hspace{0.4cm} \textrm{,} \hspace{0.6cm}
t = \eta \hspace{0.6cm} \textrm{and} \hspace{0.6cm}
{\delta\rho_{B}}_{f}(\xi,t)=\psi(\Omega,\eta)
\label{introe}
\end{equation}
where $\xi_{0}$ is an arbitrary constant.
From (\ref{introe}) we have the operators:
\begin{equation}
{\frac{\partial}{\partial \xi}}
={\frac{1}{h(\Omega,\eta)}}{\frac{\partial}{\partial \Omega}}
\hspace{0.6cm}
\textrm{and} \hspace{0.6cm}
{\frac{\partial}{\partial t}}={\frac{\partial}{\partial \eta}}-
{\frac{\beta}{h(\Omega,\eta)}}\,
\psi {\frac{\partial}{\partial \Omega}}
-{\frac{\alpha}{h(\Omega,\eta)}}\,{\frac{\partial}{\partial \Omega}}
\label{introepers}
\end{equation}
where the function $h(\Omega,\eta)$ is given by:
\begin{equation}
h(\Omega,\eta)=1+\beta \int_{-\infty}^{\eta} \Big[{\frac{\partial}
{\partial \Omega}}
\psi(\Omega,\eta')\Big] \,\, d\eta'   
\label{introepersfuncg}
\end{equation}
The equation (\ref{roegeral}) rewritten in terms of (\ref{introepers}) and
(\ref{introepersfuncg}) is:
\begin{equation}
h(\Omega,\eta)={\frac{1}{\Gamma\,\psi}}
{\frac{\partial}{\partial \Omega}}{\frac{\partial \psi}{\partial \eta}}   
\label{roealmostint}
\end{equation}
From (\ref{introepersfuncg}) we have:
\begin{equation}
{\frac{\partial h}{\partial \eta}}=\beta\,{\frac{\partial \psi}{\partial
\Omega}}   
\label{delhdeleta}
\end{equation}
Finally, inserting (\ref{roealmostint}) in (\ref{delhdeleta}) we arrive
at the following equation:
\begin{equation}
\psi\,{\frac{\partial^{2}}{\partial \eta^{2}}}\,{\frac{\partial \psi}
{\partial \Omega}}
-{\frac{\partial \psi}{\partial \eta}}{\frac{\partial}
{\partial \Omega}}{\frac{\partial \psi}{\partial \eta}}-(\beta\Gamma)
(\psi)^{2}\,{\frac{\partial \psi}{\partial \Omega}}=0
\label{roeintegravel}
\end{equation}
which is the ROE equation (\ref{roegeral}) rewritten in a integrable form. To
solve
(\ref{roeintegravel}) we apply the hyperbolic tangent function method as
described in \cite{w2,weset,tangh} and find the following exact solutions:
\begin{equation}
\psi_{I}(\Omega,\eta)=-{\frac{6\gamma^{2}\lambda^{2}}{\beta\Gamma}}sech^{2}
\Big[\lambda(\Omega-\gamma\eta) \Big]
\hspace{0.7cm} \textrm{or} \hspace{0.7cm}
\psi_{II}(\Omega,\eta)={\frac{4\gamma^{2}\lambda^{2}}{\beta\Gamma}}
+\psi_{I}(\Omega,\eta)   
\label{roesols}
\end{equation}
The parameters $\lambda$, which is the inverse of the width and $\gamma$,
the speed, are integration constants
and are free to be chosen.  The negative sign in the $-sech^{2}[\dots]$
function in (\ref{roesols}) describes a rarefaction pulse.
Such negative sign is due the condition $(\alpha\Gamma)>0$.

Considering $\psi_{I}$ from (\ref{roesols}) in (\ref{introe}) we obtain the
following parametric solution of (\ref{roegeral}):
\begin{equation}
{{\delta\rho_{B}}_{f}}(\xi,t)=-{\frac{6\gamma^{2}\lambda^{2}}
{\beta\Gamma}}sech^{2}
\Big[\lambda\Big(\Omega-\gamma t\Big) \Big]
\label{roesolsexacta-ap}
\end{equation}
with
\begin{equation}
\xi=\Omega +\alpha\,t +\xi_{0}
+{\frac{6\gamma\lambda}{\beta\Gamma}}\Big\{tanh\Big[\lambda
\Big(\Omega-\gamma t\Big) \Big]
-1\Big\}   
\label{introeagain-ap}
\end{equation}

As previously mentioned, the last two expressions are (\ref{roesolsexacta}) and (\ref{introeagain}), respectively.

We do not consider $\psi_{II}$ of (\ref{roesols}) as solution of
(\ref{roegeral}). The reason is to avoid the divergence due the constant
term of $\psi_{II}$ in the integral present in (\ref{introe}):
$$
\beta \int_{-\infty}^{\eta}{\frac{4\gamma^{2}\lambda^{2}}{\beta\Gamma}} \,
d\eta' \rightarrow \infty  
$$

\begin{acknowledgments}

\vskip0.5cm

This work was partially supported by the Brazilian funding agencies CAPES, CNPq and
FAPESP (contract 2012/98445-4).

\end{acknowledgments}

\end{document}